\documentclass[reprint,amsmath,amssymb,aps,pra,superscriptaddress]{revtex4-1}
\usepackage{amsmath,epsfig,wrapfig,graphicx,color,bm}

\begin{document}
\title{Zero-energy excitation in the classical kagome antiferromagnet NaBa$_{2}$Mn$_{3}$F$_{11}$}

\author{Shohei~Hayashida}
\affiliation{Institute for Solid State Physics, The University of Tokyo, Chiba 277-8581, Japan}
\author{Hajime~Ishikawa}
\affiliation{Institute for Solid State Physics, The University of Tokyo, Chiba 277-8581, Japan}
\author{Yoshihiko~Okamoto}
\affiliation{Department of Applied Physics, Nagoya University, Nagoya 464-8603, Japan}
\author{Tsuyoshi~Okubo}
\affiliation{Department of Physics, The University of Tokyo, Tokyo 113-0033, Japan}
\author{Zenji~Hiroi}
\affiliation{Institute for Solid State Physics, The University of Tokyo, Chiba 277-8581, Japan}
\author{G{\o}ran J.~Nilsen}
\affiliation{ISIS Neutron and Muon Source, Science and Technology Facilities Council, Didcot
 OX11 0QX, United Kingdom}
\author{Hannu~Mutka}
\affiliation{Institute Laue Langevin, 71 avenue des Martyrs, 38042 Grenoble, France}
\author{Takatsugu~Masuda}
\affiliation{Institute for Solid State Physics, The University of Tokyo, Chiba 277-8581, Japan}

\date{\today}

\begin{abstract}
We performed inelastic neutron scattering measurements on a polycrystalline sample of
a classical kagome antiferromagnet NaBa$_{2}$Mn$_{3}$F$_{11}$ to investigate the possibility of
a dispersionless zero-energy excitation associated with rotation of spins along the chains.
The observed spectra indeed exhibit such an excitation with strong intensity at low energy,
as well as dispersive excitations with weak intensity at high energy.
Combining the measurements with calculations from linear spin-wave theory reveals 
that NaBa$_{2}$Mn$_{3}$F$_{11}$ is a good realization of the classical kagome antiferromagnet which exhibits a dispersionless mode lifted by the magnetic dipole-dipole interaction. 
\end{abstract}

\maketitle

\section{Introduction}
Geometrical frustration has been extensively studied in terms of both its theoretical and experimental aspects in condensed-matter physics~\cite{Lacroix2011,Diep2013}.
Frustrated systems retain macroscopic degeneracy even at low temperatures, providing diverse and exotic spin states~\cite{Balents2010}.
One of the remarkable phenomena is localization of spin-wave excitations.
For classical spin systems, i.e., continuous spins, magnetic structures at the ground state are largely degenerate due to the frustration.
The degenerate magnetic structures allow a continuous rearrangement of the spins with no energy cost, generating a dispersionless mode in the spin-wave excitation spectrum.
This means that the spin wave is localized in momentum space.
Away from geometrically frustrated magnets, dispersionless bands have attracted great interest.
They have been proposed to be key to a variety of exotic phenomena, including the unconventional topological orders in fermionic systems~\cite{Parameswaran2013,Bergholtz2013,Derzhko2015}
and the magnon Hall effect in ferromagnetic insulators~\cite{Ohgushi2000,Ideue2012,Zhang2013,Chisnell2015}.

\begin{figure}[tbp]
\includegraphics[scale=1]{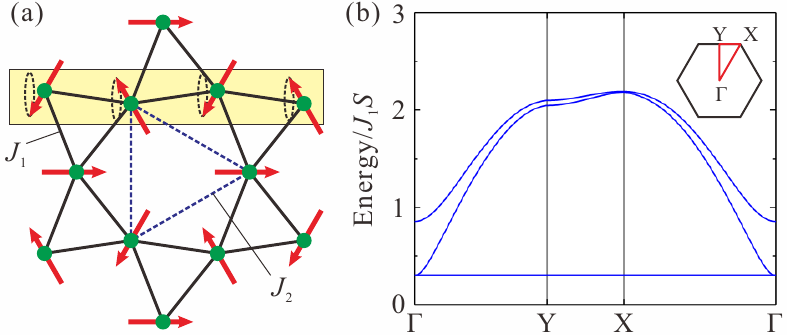}
\caption{(a) Magnetic structure having ${\bm k}={\bm 0}$ of NaBa$_{2}$Mn$_{3}$F$_{11}$. 
The red arrows represent directions of spins.
Dashed loops illustrate the zero-energy mode as described in the text.
Solid and dashed lines are the nearest-neighbor and second-neighbor paths, respectively.
(b) Spin-wave excitation having the nearest-neighbor exchange interaction $J_{1}$ and MDD
interaction $J_{\rm MDD}=J_{1}/100$. The energy is normalized by the magnitude of the interaction $J_{1}$ and the spin $S$.}
\label{fig1}
\end{figure}

The classical kagome antiferromagnet is the prototypical system for a dispersionless mode in the spin-wave excitations.
It has an infinite degeneracy of $120^{\circ}$ structures in the ground state~\cite{Harris1992,Chalker1992,Huse1992,Reimers1993}.
This degeneracy allows a continuous change of the spin arrangement.
For example, in the case of the so-called ${\bm q}=0$ structure, two spins in a triangle can rotate about the direction of the rest of the spins, while retaining the $120^{\circ}$ configuration.
There is therefore no energy cost associated with the excitations.
The rotating spins form a chain, as illustrated in Fig.~\ref{fig1}(a).
A set of spins in each chain are excited independently on spins in different chains, meaning that the excitation is localized in the chain. 
The spin rotation with no energy cost is a localized mode, namely a {\it zero-energy mode}~\cite{Harris1992,Chalker1992,Ritchey1993}.
This produces zero-energy lines in the magnetic Brillouin zone.

In real kagome antiferromagnets, the macroscopic degeneracy of the 120$^{\circ}$ structures is solved by
some types of magnetic anisotropy such as the Dzyaloshinskii-Moriya interaction, single-ion anisotropy, and magnetic dipole-dipole (MDD) interaction. 
Then, the zero-energy mode becomes visible as an excited state lifted by those anisotropies.
In potassium iron jarosite KFe$_{3}$(OH)$_{6}$(SO$_{4}$)$_{2}$, 
an excitation at 7 meV was found to be a zero-energy mode lifted by the Dzyaloshinskii-Moriya interaction through linear spin-wave calculations~\cite{Matan2006}.
It is, however, made dispersive by the second-neighbor exchange interaction, which couples the chains.
To our knowledge, the zero-energy mode in the kagome antiferromagnet has been reported only in
KFe$_{3}$(OH)$_{6}$(SO$_{4}$)$_{2}$.
Further study in different materials is thus important.

\begin{figure*}[htbp]
\includegraphics[scale=1]{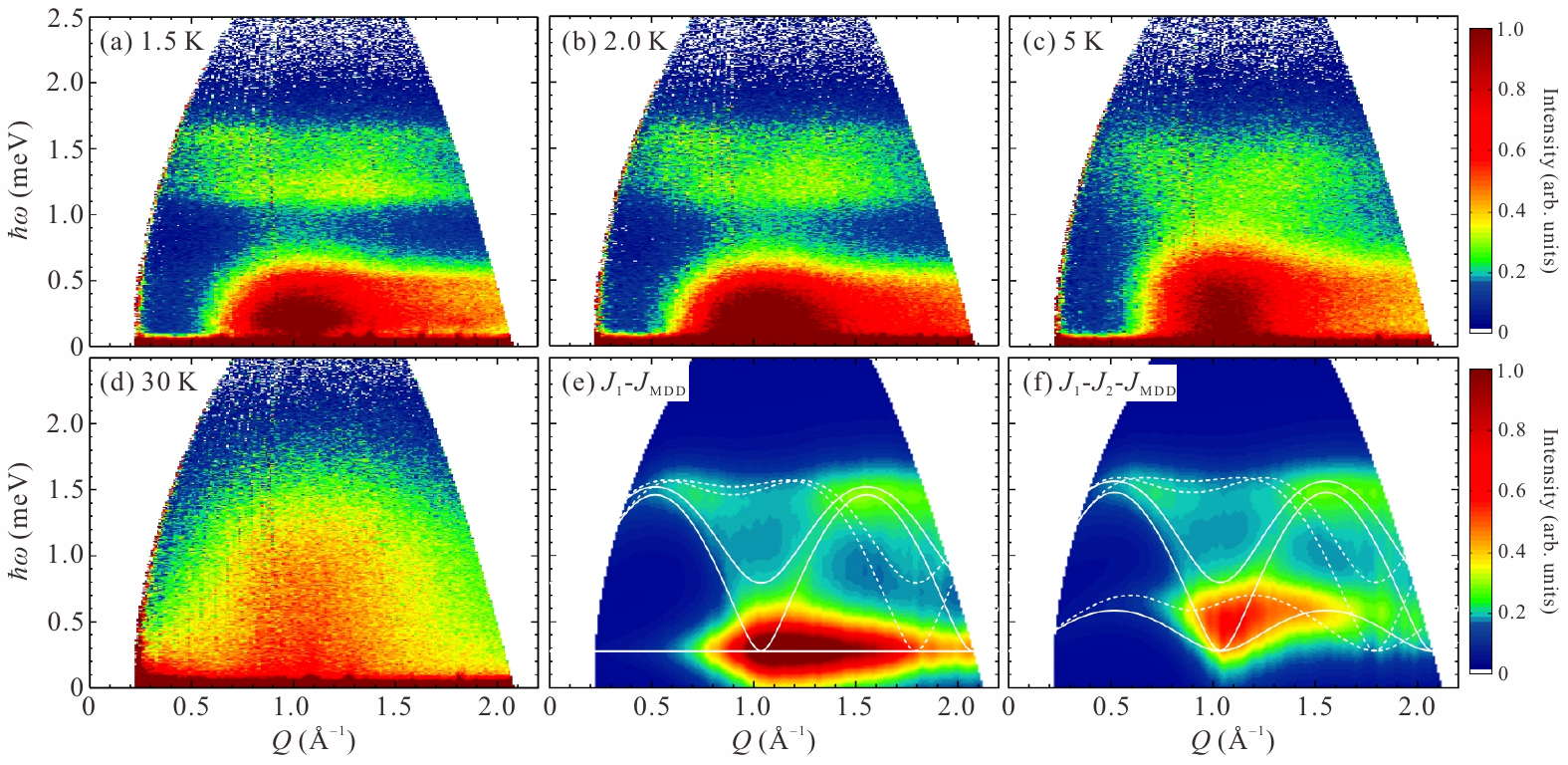}
\caption{INS spectra of NaBa$_{2}$Mn$_{3}$F$_{11}$ at (a) $1.5$~K, (b) $2.0$~K, (c) $5$~K, and
(d) 30 K. The incident neutron energy is $E_{\rm i}=3.1$ meV.
Calculated spin-wave spectra with (e) $J_{1}= 0.28$~meV, $J_{2}= 0$ 
 and $J_{{\rm MDD}}= 4.9$~$\mu$eV and with (f) $J_{1}= 0.27$~meV, $J_{2}=J_{1}/10$ 
 and $J_{{\rm MDD}}= 4.9$~$\mu$eV. Solid and dashed curves in (e) and (f) 
are spin-wave dispersions along $[1~0~0]$ and $[1~1~0]$ directions, respectively.}
\label{fig2}
\end{figure*}

Our target compound is a classical kagome antiferromagnet NaBa$_{2}$Mn$_{3}$F$_{11}$.
This compound crystallizes in a hexagonal structure with the space group $R\bar{3}c$~\cite{JSSC98}.
The Mn$^{2+}$ ions carry spin $S=5/2$, and MnF$_{7}$ pentagonal bipyramids form a 
kagome lattice in the crystallographic $ab$ plane.
Thermodynamic measurements exhibit a Curie-Weiss temperature of $\theta_{\rm CW}=-32$~K and an antiferromagnetic transition at $T_{{\rm N}}=2$~K~\cite{Ishikawa2014}.
Neutron powder diffraction identified that the basic magnetic structure is the 
$120^{\circ}$ structure with the magnetic propagation vector ${\bm k}_{0}=(0,0,0)$ 
in Fig.~\ref{fig1}(a), and it is modulated incommensurately~\cite{Hayashida2018}.
A calculation of the ground state including the nearest-neighbor antiferromagnetic interaction $J_{1}$, 
the second-neighbor antiferromagnetic interaction $J_{2}$, and
a MDD interaction $J_{\rm MDD}$ showed that the identified $120^{\circ}$ structure was selected by the MDD interaction.

Theoretical studies have shown that the classical kagome antiferromagnet with the MDD interaction 
has the zero-energy mode as its lowest excited state~\cite{Maksymenko2015,Maksymenko2017}, 
as shown in Fig.~\ref{fig1}(b).
Since the flatness of the excitation is robust against long-range MDD interactions~\cite{Maksymenko2017}, 
the observation of a dispersionless zero-energy mode is expected in NaBa$_{2}$Mn$_{3}$F$_{11}$. 
In the present paper, we investigate the zero-energy mode in NaBa$_{2}$Mn$_{3}$F$_{11}$ through 
a combination of inelastic neutron scattering (INS) experiments and spin-wave calculations.
The observed energy of the dispersionless mode matches the anisotropy gap originating from the MDD interaction.

\section{Experimental details}\label{sec2}
A 19~g polycrystalline sample of NaBa$_{2}$Mn$_{3}$F$_{11}$ was prepared by a solid-state reaction method~\cite{Ishikawa2014}.
We loaded the sample in a copper cell, which was installed in a $^{4}$He cryostat which achieves $1.5$~K.
The INS experiment was performed at the cold-neutron time-of-flight (TOF) 
spectrometer IN6 at the Institut Laue-Langevin (ILL) in Grenoble, France.
The energy of the incident neutron beam was $E_{\rm i}=3.1$~meV, yielding a Gaussian energy 
resolution of $\Delta E= 0.07$ meV at the elastic position.  
A preliminary experiment was performed at the thermal-neutron TOF spectrometer IN4C 
at the same institution to measure the magnetic excitations up to $6$~meV.
The absence of magnetic excitations above $2.5$~meV was confirmed.
These INS spectra are shown in the Appendix.

\section{Experimental results}\label{sec3}
\begin{figure}[tbp]
\includegraphics[scale=1]{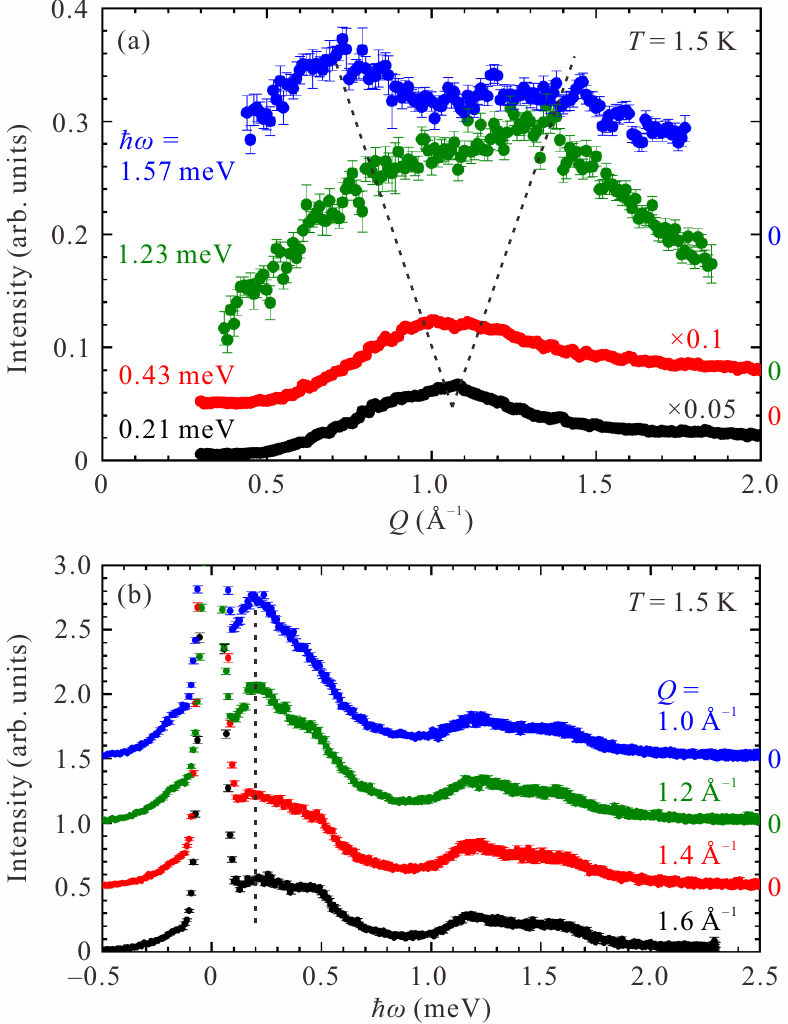}
\caption{(a) Constant-energy cuts at $\hbar\omega= 0.21$, $0.43$, $1.23$, and $1.57$~meV 
and at $1.5$~K. The data are integrated in the range of $\hbar\omega\pm 0.05$~meV. 
(b) Constant-$Q$ cuts at $Q= 1.0$, $1.2$, $1.4$, and $1.6$~{\AA}$^{-1}$ and at $1.5$~K. The data are integrated
in the range of $Q\pm0.2$~{\AA}$^{-1}$.
The dashed lines are guides for eyes.}
\label{fig3}
\end{figure}

In the magnetic ordered state at $1.5$~K, excitations with strong intensity at $0.2$~meV
and weak intensity at $1.5$~meV are observed, as shown in Fig.~\ref{fig2}(a).
The center of mass in $Q$ of the strong spectral weight is at $Q\sim 1$~{\AA}$^{-1}$,
which is observed as a broad peak in the constant energy cut at $0.21$~meV in Fig.~\ref{fig3}(a).
The $Q$ position of the peak maximum coincides with the strongest magnetic Bragg reflection $(1~0~1)$.
The peak splits into two peaks with increasing energy, as indicated by the dashed lines in Fig.~\ref{fig3}(a).
This implies that a spin-wave excitation disperses from $(1~0~1)$.
In a series of constant-$Q$ cuts in Fig.~\ref{fig3}(b), the strong peak is identified at $0.21$~meV.
This peak position is attributed to an anisotropy energy gap, which is compatible with the scale of the ordering temperature $T_{\rm N}=2$~K.
It is notable that this peak does not shift on varying $Q$, indicating that the excitation at $0.21$~meV is dispersionless.
As the MDD interaction creates a gap in the excitation spectrum, with the zero-energy mode
immediately above it~\cite{Maksymenko2015,Maksymenko2017},
the excitation at $0.21$~meV is expected to be the zero-energy mode. 
The upper boundary of the spin-wave dispersion is evaluated to be $1.57$~meV.
Weak intensity is observed up to $2.5$~meV, and it is well fit by a Lorentzian tail of the peak 
at $1.57$~meV.
Note that the intensity remains even at $60$~mK, which was found in the spectrum in the IN4C
 [see Fig.~\ref{fig7}(a) in the Appendix].
This implies that the spin-waves are damped by persistent spin-fluctuations much below $T_{\rm N}=2$~K.

\begin{figure}[tbp]
\includegraphics[scale=1]{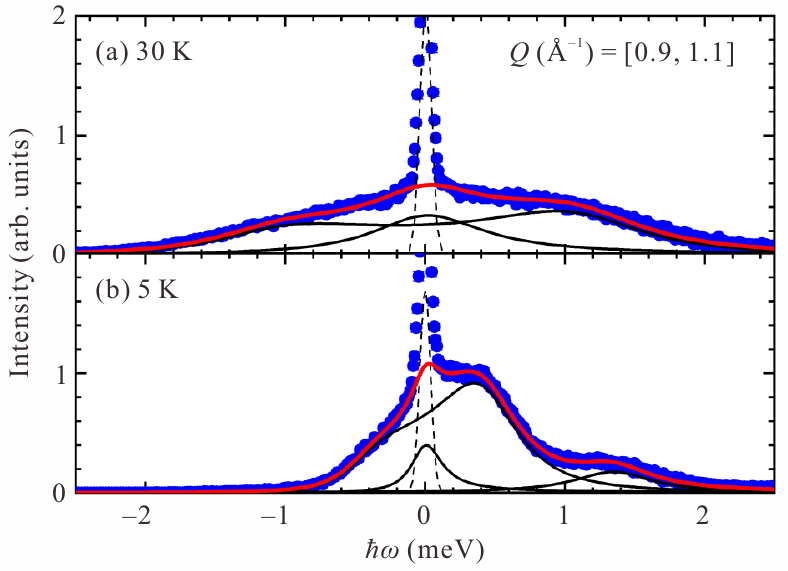}
\caption{Constant-$Q$ scans at $Q=1$ {\AA}$^{-1}$ for (a) $30$~K and (b) $5$~K. 
The spectra are integrated in the range of $Q=[0.9, 1.1]$.
The dashed curves are elastic lines fitted by Gaussian functions.
The solid curves are fits to data by Lorentzian functions. 
The red curves are the sum of the black solid curves.}
\label{fig4}
\end{figure}

In the paramagnetic state at $30$~K, strong magnetic diffuse scattering indicative of short-range correlations is observed, as shown in Fig.~\ref{fig2}(d).
This means that the spin correlation develops at much higher temperature than the transition
temperature $T_{\rm N}=2$ K.
In Figs.~\ref{fig2}(b) and \ref{fig2}(c) still above $T_{\rm N}$, the diffuse scattering is suppressed and the spectra are split into low 
and high energy parts with decreasing temperature.
In other words, the magnetic excitation becomes structured upon approaching the transition temperature, 
owing to the further development of longer ranged spin correlations.

The temperature evolution of constant-$Q$ cuts at $Q=1$~{\AA}$^{-1}$ are shown in 
Figs.~\ref{fig4}(a) and \ref{fig4}(b).
The quasielastic scattering spectra in the paramagnetic state are fitted by the dynamical 
structure factor $S(Q,\omega,T)$~\cite{Fak2008,Krimmel2009} with an exponential spin 
relaxation in the form of a Lorentzian-shaped function as follows:
\begin{eqnarray}
S(Q,\omega,T)=\frac{1}{1-e^{-\hbar\omega/k_{\rm B} T}}
\frac{\chi_{0}(Q,T)\omega\Gamma(Q,T)}{\omega^{2}+\Gamma(Q,T)^{2}},
\end{eqnarray}
where the first term represents the detailed balance factor accounting for thermal 
population of the excited state, and $\Gamma$ is the line width.
$\chi_{0}(Q,T)$ is the static susceptibility.
The INS spectra are fitted by an additional damped harmonic oscillator (DHO), considering
detailed balance and corresponding to the double Lorentzian function~\cite{Krimmel2009,FAK19971107} represented as follows:
\begin{eqnarray}
&&S(Q,\omega,T) \nonumber \\
&&=\frac{1}{1-e^{-\hbar\omega/k_{\rm B} T}}\frac{A_{\rm DHO}(Q,T)\omega\Gamma (Q,T)}{\left(\omega^{2}-\omega_{\rm DHO}^{2}\right)^{2}+\left(\omega\Gamma(Q,T)\right)^{2}},
\end{eqnarray}
where $A_{{\rm DHO}}(Q,T)$ is the oscillator strength,  and $\omega_{{\rm DHO}}$ is the eigenfrequency.

The spectrum at $30$ K, which is close to the Curie-Weiss temperature $\theta_{{\rm CW}}= -32.3$~K~\cite{Ishikawa2014},  is well described by a quasielastic Lorentzian and an inelastic double Lorentzian as shown in Fig.~\ref{fig4}(a). 
While the quasielastic spectrum coming from heavily damped spin waves in the paramagnetic state is expected, the inelastic feature centered at $1.3$~meV is more surprising.
This means that the spin correlations with respect to time develops significantly even at $30$~K. 
The quasielastic spectrum is suppressed and the inelastic spectrum is enhanced at $5$~K, as indicated in Fig.~\ref{fig4}(b), i.e., longer ranged correlations are present.
This means that the paramagnetic response transfers to the inelastic 
as the spin correlations further develop on approaching the transition temperature.

\section{Analysis}\label{sec4}
To identify the magnetic model of NaBa$_{2}$Mn$_{3}$F$_{11}$, we calculate
the spin-wave excitation spectra in linear spin-wave theory.
We assume that the $120^{\circ}$ structure mainly contributes to the spectra, and the incommensurate
modulation is not considered for simplicity. 
We consider the following Hamiltonian:
\begin{eqnarray}
&{\mathcal H}&=\sum_{{\rm n.n.}}J_{1}{\bm S}_{i}\cdot{\bm S}_{j}
+\sum_{{\rm n.n.n.}}J_{2}{\bm S}_{i}\cdot{\bm S}_{j}
 \nonumber \\
&+&\sum_{{\rm n.n.}}\frac{\mu_{0}}{4\pi}\frac{(g\mu_{{\rm B}})^{2}}{|{\bm r}_{ij}|^{3}}\left[{\bm S}_{i}\cdot{\bm S}_{j}
-3\frac{\left({\bm S}_{i}\cdot{\bm r}_{ij}\right)\left({\bm S}_{j}\cdot{\bm r}_{ij}\right)}
{|{\bm r}_{ij}|^{2}}\right], \nonumber \\
\end{eqnarray}
where $J_{1}$ and $J_{2}$ are the nearest- and second-neighbor exchange paths as shown in Fig.~\ref{fig1}(a).
These are both fixed to be antiferromagnetic to realize the $120^{\circ}$ structure with ${\bm k}_{0}=(0,0,0)$~\cite{Hayashida2018}.
The third term is the nearest-neighbor MDD interaction with the bond vector ${\bm r}_{ij}$ between the 
spins.
The strength of the nearest neighbor MDD interaction $J_{{\rm MDD}}$ is fixed:
$J_{{\rm MDD}}={\mu_{0}(g\mu_{{\rm B}})^{2}}/{4\pi}{r_{{\rm n.n.}}^{3}}$,
where $r_{\rm n.n.}$ is the distance of the nearest neighbor path.
From this, $J_{\rm MDD}$ is estimated to be $4.9$~$\mu$eV.
Note that we ignore further-neighboring MDD interactions in this calculation because 
they insignificantly affect the spin-wave spectrum~\cite{Maksymenko2017}. 

The spin-wave dispersion is calculated based on the linear spin-wave  theory using 
the Holstein-Primakoff formalism~\cite{Petit2011}.
The spin-wave spectra were calculated and then powder averaged using the SpinW
package~\cite{spinw}.
The powder averaged spectra were convoluted by a Gaussian function with 
a full-width half-maximum (FWHM) $\Delta Q= 0.03$~{\AA}$^{-1}$ and a Lorentzian function with
a FWHM $\Delta E= 0.17$~meV.
$\Delta Q=0.03$~{\AA}$^{-1}$ is evaluated from a Gaussian fit of the Bragg peaks.
Since the spin waves are damped by persistent spin-fluctuation, we used a Lorentzian function along the energy.
$\Delta E=0.17$~meV is optimized by a chi-squared analysis of observed and calculated constant-$Q$ cuts at $Q= 1$~{\AA}$^{-1}$.

\begin{figure}[tbp]
\includegraphics[scale=1]{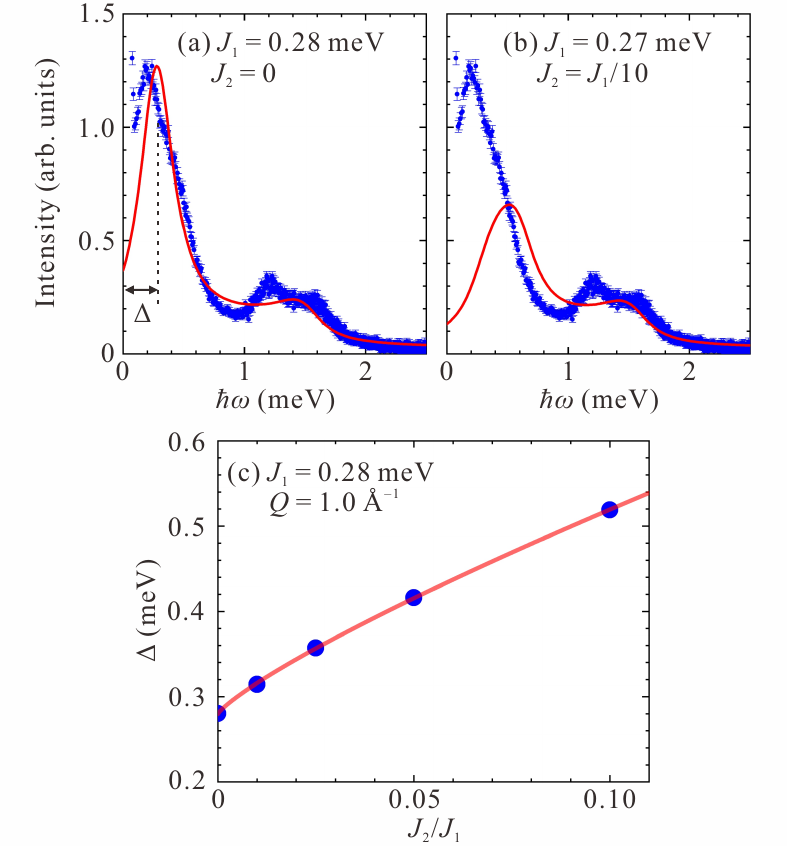}
\caption{Comparison of constant-$Q$ cut at $Q= 1$~{\AA}$^{-1}$ between the
experimental result at $T=1.5$~K and calculations. 
Blue marks are the experimental data.
The red curves are the calculation in which the interactions are (a) $J_{1}= 0.28$~meV, 
$J_{2}= 0$~meV, and $J_{{\rm MDD}}= 4.9$~$\mu$eV, and (b) $J_{1}= 0.27$~meV, 
$J_{2}=J_{1}/10$, and $J_{{\rm MDD}}= 4.9$~$\mu$eV.
The spectra are integrated in the range of $Q=[0.9, 1.1]$.
$\Delta$ indicates the peak position of the low energy at $Q= 1$~{\AA}$^{-1}$.
(c) $J_{2}$ variation of $\Delta$ in the calculation with $J_{1}= 0.28$~meV and $J_{{\rm MDD}}= 4.9$~$\mu$eV.
The red curve is a guide for eye.}
\label{fig5}
\end{figure}

We calculate the full powder-averaged spectra in two cases: (i) for the $J_{1}$-$J_{{\rm MDD}}$ and (ii) for the $J_{1}$-$J_{2}$-$J_{{\rm MDD}}$ 
models.
In the latter model, we set $J_{2}=J_{1}/10$ for simplicity.
The strength of $J_{1}$ is evaluated by setting the upper boundary of the spectrum as $1.57$~meV, and
is found to be $0.28$~meV for the $J_{1}$-$J_{{\rm MDD}}$ model and $0.27$~meV for the
$J_{1}$-$J_{2}$-$J_{{\rm MDD}}$ model. 
From $(J_{1},J_{2})=(0.28~{\rm meV},0~{\rm meV})$ and $(0.27~{\rm meV},0.027~{\rm meV})$, the Curie-Weiss temperatures $\theta_{\rm CW}=-(z_{1}J_{1}+z_{2}J_{2})S(S+1)/3k_{\rm B}$ are estimated 
to be $-37.9$~K and $-38.4$~K, where $S=5/2$ is the Mn$^{2+}$ spin,  $z_{1}=4$ and $z_{2}=2$ are the coordination numbers of the nearest- and second-neighboring paths, and $k_{\rm B}$ is the Boltzmann constant.
These values are consistent with the $\theta_{\rm CW}=-32.3$~K evaluated by the magnetic 
susceptibility~\cite{Ishikawa2014}. 

The calculated dispersions show three modes as displayed in Fig.~\ref{fig2}(e) for the $J_{1}$-$J_{{\rm MDD}}$ model
and Fig.~\ref{fig2}(f) for the $J_{1}$-$J_{2}$-$J_{{\rm MDD}}$ model.
The excitation at the lowest energy in the former model is dispersionless, and 
it corresponds to the zero-energy mode lifted to finite energy by the MDD interaction.
This result is consistent with previous theoretical studies~\cite{Maksymenko2015,Maksymenko2017}.
The zero-energy mode becomes dispersive when $J_2$ is included as shown in Fig.~\ref{fig2}(f).
Comparing with the experiment, the observed spectrum at $0.21$~meV is more similar to the calculated spectrum in the $J_1$-$J_{\rm MDD}$ model rather than the one in the $J_1$-$J_2$-$J_{\rm MDD}$ model.
This means that the observed spectrum at $0.21$~meV is probably the dispersionless excitation lifted by the MDD interaction, and $J_2$ is negligible compared with $J_1$ in NaBa$_{2}$Mn$_{3}$F$_{11}$.
The observed spectrum around $0.2$~meV and at $1.5$~K is broader than the calculated spectrum of the $J_1$-$J_{\rm MDD}$ model, implying that strong spin-fluctuations and/or disorder in the system
remain at $1.5$~K.

The calculation and experiment at constant-$Q$ cut are shown in Figs.~\ref{fig5}(a) and \ref{fig5}(b).
There are two structures at $0.28$ and $1.40$~meV in the $J_1$-$J_{\rm MDD}$ model and 
at $0.50$ and $1.42$~meV in the $J_1$-$J_2$-$J_{\rm MDD}$ model.
In the $J_1$-$J_{\rm MDD}$ model, the calculated spectrum semi-quantitatively reproduces 
the anisotropy gap of $0.21$~meV, which means that the MDD interaction is the main contributor to 
the magnetic anisotropy.
In contrast, the $J_1$-$J_2$-$J_{\rm MDD}$ model exhibits a broadened low energy peak shifted to the high energy, which no longer matches to the position of the anisotropy gap, as shown in Fig.~\ref{fig5}(b).
$J_{2}/J_{1}$ dependence of the peak position of the low energy at $Q = 1$~{\AA}$^{-1}$, $\Delta$,  
is shown in Fig.~\ref{fig5}(c). 
$\Delta$ increases with $J_{2}/J_{1}$ and deviates from the low-energy gap observed experimentally.
These results reinforce that the observed zero-energy excitation at $0.2$~meV is mainly lifted by the MDD interaction, and the second-neighbor interaction $J_2$ is negligible in NaBa$_{2}$Mn$_{3}$F$_{11}$.
We have also confirmed that inclusion of the third-neighbor interaction in the $ab$ plane $J_{3}$ makes the anisotropy gap shift to higher energy in any combination of $J_{2}$ and $J_{3}$, as far as the combination realizes the 120$^{\circ}$ structure.

\section{Discussion}\label{sec5}
The observed dispersionless mode in NaBa$_{2}$Mn$_{3}$F$_{11}$ is a unique signature of classical kagome physics.
We found that the energy position of the dispersionless mode is reproduced solely by the MDD interaction, 
and that second-neighbor interaction $J_{2}$ is negligible.
Note that the MDD interaction is ubiquitous in every real magnet even though the qualitative behavior of 
most kagome antiferromagnet can be explained by models including only the nearest-neighbor interaction~\cite{Harris1992,Chalker1992,Huse1992,Reimers1993,Ritchey1993}.
In addition, further neighbor interactions significantly influence the dispersionless zero-energy mode.
It suppresses the continuous rearrangement of the spin with no energy cost, and makes the zero-energy mode dispersive~\cite{Harris1992}.
Therefore, we conclude that the observed dispersionless excitation is the ideal zero-energy mode in the realistic
classical kagome antiferromagnet.

The observed anisotropy gap of $0.21$~meV is $25${\%} smaller than the calculated one of $0.28$~meV.
This reduction is consistent with other systems in which the anisotropy gap in the spin-wave excitation closes upon approaching the transition temperature~\cite{Oguchi1961,Fulton1994,Wildes1998}.
The temperature dependence of the gap in antiferromagnets is known to be roughly proportional to the sublattice magnetization.
In the neutron diffraction experiment on NaBa$_{2}$Mn$_{3}$F$_{11}$,  the sublattice magnetic moment 
at $1.5$~K is $80${\%} of that at $0.25$~K~\cite{Hayashida2018}.
The reduction of the anisotropy gap is thus expected to be $20${\%} at $1.5$~K.
This value approximately coincides to the $25${\%} reduction of the gap between the experiment and calculation.
Accordingly, we conclude that the main anisotropy lifting the zero-energy mode is still the MDD interaction in spite of the reduction of the gap.

\section{Conclusion}\label{sec6}
In conclusion,  the magnetic excitations of NaBa$_{2}$Mn$_{3}$F$_{11}$ measured by inelastic neutron scattering exhibit a dispersionless excitation at $0.2$~meV. 
The calculations based on linear spin-wave theory reveals that the excitation is
described by the zero-energy mode lifted mainly by the MDD interaction. 
Thus, NaBa$_{2}$Mn$_{3}$F$_{11}$ is a unique classical kagome antiferromagnet exhibiting a truly 
dispersionless lifted zero-energy excitation.
For future work, further studies such as measurements using single-crystal
samples and more detailed spin-wave calculations are necessary to elucidate the physical origin of the 
additional rather than unusual magnetic excitations.

\section*{Acknowledgements}
Travel expenses for the inelastic neutron scattering experiments performed using IN4C
and IN6 at ILL, France, were supported by the General User
Program for Neutron Scattering Experiments, Institute for Solid State Physics,
The University of Tokyo Proposal (No.~14528), at JRR-3, Japan Atomic Energy Agency, Tokai, Japan.
This work was partially supported by KAKENHI (Grand No.~15K17701 and No.~19K03740).
S.H. was supported by the Japan Society for the Promotion of  Science through the 
Leading Graduate Schools (MERIT).

\section*{Appendix: Inelastic neutron spectra measured at the IN4C spectrometer}
\begin{figure*}[htbp]
\includegraphics[scale=1]{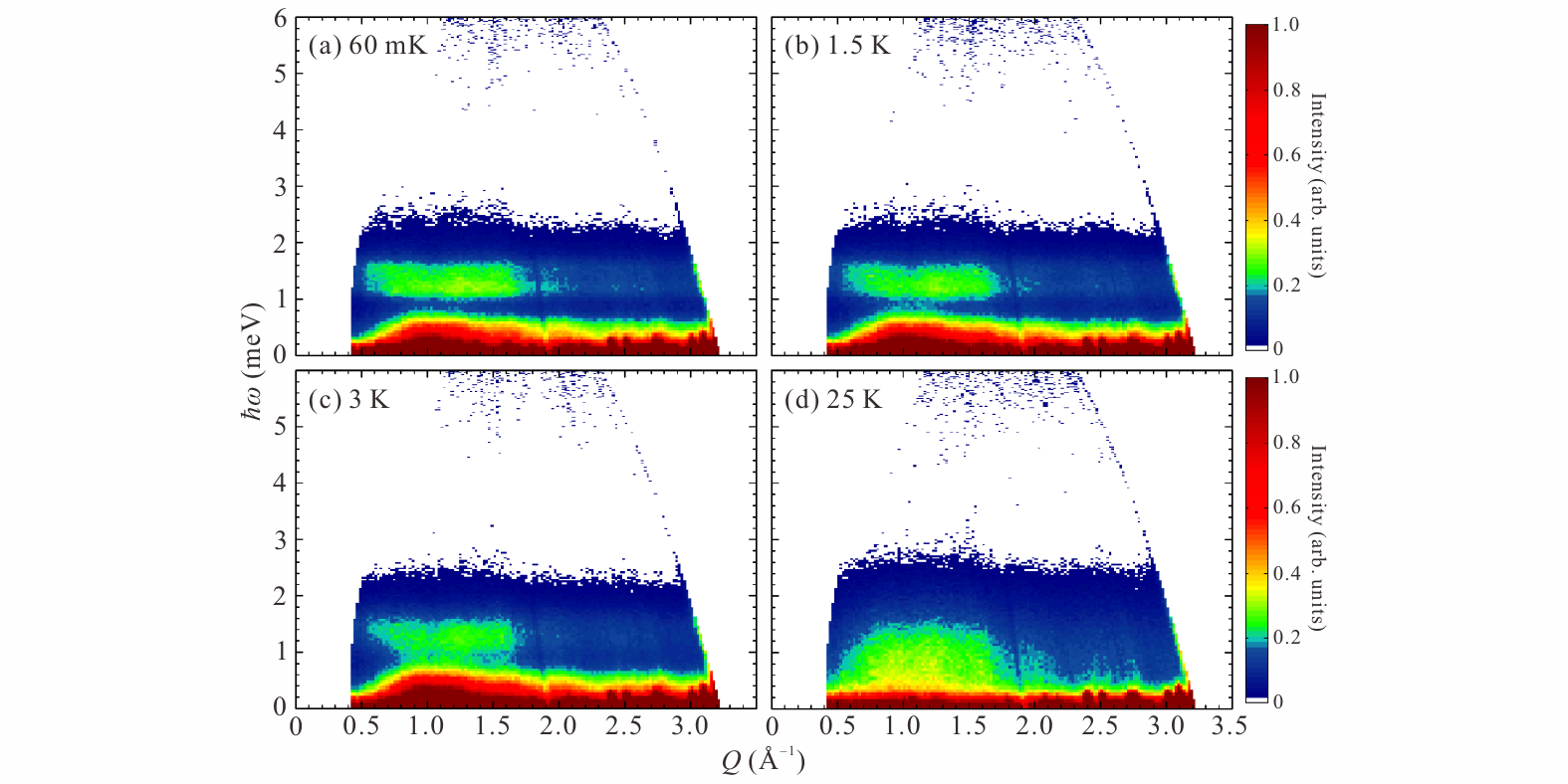}
\caption{INS spectra measured at the IN4C spectrometer at (a) 60~mK, (b) 1.5~K, (c) 3~K, and
(d) 25 K.
The incident neutron energy is $E_{\rm i}=7.1$ meV.
}
\label{fig7}
\end{figure*}

In the INS experiment performed at the IN4C spectrometer, the same sample as measured by the IN6 spectrometer was used, and it was installed in a dilution refrigerator achieving $60$~mK.
The energy of the incident neutron beam was $7.1$~meV, yielding a Gaussian energy resolution of
$\Delta E=0.31$~meV at the elastic position.
INS spectra measured by the IN4C spectrometer are displayed in Fig.~\ref{fig7}.
Below $5$~K, there is an excitation at $1.5$~meV in agreement with the weak intensity observed in 
Figs.~\ref{fig2}(a)--\ref{fig2}(c).
The strong intensity is also observed below $1$~meV but its structure is unclear because of overlap with elastic incoherent scattering.  
 Magnetic diffuse quasielastic scattering is also observed at $25$~K, which is totally consistent with the spectrum at $30$~K in Fig.~\ref{fig2}(d).
At all temperatures, no intensity is observed above $2.5$~meV.

\end{document}